\documentstyle[aps,pre,manuscript]{revtex}

\begin{document}

\draft

\title{The generalized Kramers' theory for an external noise driven bath}

\author{Jyotipratim Ray Chaudhuri, Suman Kumar Banik and Deb Shankar Ray
\footnote{e-mail : pcdsr@mahendra.iacs.res.in} }

\address{Indian Association for the Cultivation of Science, Jadavpur,
Calcutta 700 032, India.}

\date{\today}

\maketitle

\begin{abstract}
We consider a system-reservoir model where the reservoir is modulated by an
external noise. Both the internal noise of the reservoir and the external
noise are stationary, Gaussian and are characterized by arbitrary decaying
correlation functions. Based on a relation between the dissipation of the
system and the response function of the reservoir driven by external noise 
we derive the generalized Kramers' rate for this nonequilibrium open system.
\end{abstract}

\vspace{0.5cm}


\section{ Introduction 
\label{sd1} 
}

More than half a century ago Kramers proposed a diffusion model for
chemical reactions in terms of the theory of Brownian motion in phase space
\cite{rd1}.
Since then the model and several of its variants have been ubiquitious in
many areas of physics, chemistry and biology for understanding the nature
of activated processes in 
classical \cite{rd2,rd3,rd4,rd5,rd6,rd7,rd8,rd9},
quantum and semiclassical \cite{rd10,rd11,rd12,rd13,rd14,rd15}
systems, in general. These
have become the subject of several reviews \cite{rd3,rd16,rd17,rd18} 
and monograph \cite{rd19} in the recent past.

In the majority of these treatments one is essentially concerned with an
equilibrium thermal bath at a finite temperature which stimulates the 
reaction co-ordinate to cross the activation energy barrier. The inherent
noise of the medium is of internal origin. This implies that the dissipative 
force which the system experiences in course of its motion in the medium
and the stochastic force acting on the system as a result of random impact
from the constituents of the medium arise from a common mechanism. From a 
microscopic point of view the system-reservoir Hamiltonian description
\cite{rd20,rd21,rd22,rd23}
developed over the decades suggests that the coupling of the system and
the reservoir co-ordinates determines both the noise and the dissipative 
terms in the Langevin equation describing the motion of the system. It is 
therefore not difficult to anticipate that these two entities get related
through a fluctuation-dissipation relation \cite{rd24} 
( these systems are sometimes classified as thermodynamically closed system 
in contrast to the systems driven by external noise in 
nonequilibrium statistical mechanics \cite{rd25} ). However,
when the reservoir is modulated by an external noise it is likely that it
induces fluctuations in the polarization of the reservoir. These fluctuations 
in turn may drive the system in addition to usual internal noise of the
reservoir. Since the polarization fluctuations of the reservoir crucially
depend on its response functions, one can further envisage a connection 
between the dissipation of the system and the response function of the 
reservoir due to the external noise from a microscopic point of view.

In the present paper we explore this connection in the context of activated
rate processes when the reservoir is modulated by an external noise. 
Specifically our object here is to calculate the generalized Kramers' rate
for the steady state of a nonequilibrium open system. Both the
internal and the external noises are
Gaussian, stationary and are characterized by arbitrary decaying correlation
functions. While the internal noise of the reservoir is thermal, the external 
noise may be of
thermal or non-thermal type. We consider the stochastic motion to be spatial
diffusion limited and calculate the rate of escape in the intermediate to
strong damping regime. We further mention that nonequilibrium,
non-thermal systems have also been investigated phenomenologically by
a number of workers in several other contexts, e.g., for examining the role
of color noise in stationary probabilities \cite{rd26}, 
properties of nonlinear systems \cite{rd27}, nature of crossover \cite{rd28}, 
rate of diffusion limited coagulation processes \cite{rd29}, 
effect of monochromatic noise \cite{rd30}, etc. While these treatments
concern direct driving of the system by an external noise the present
consideration is based on modulation of the bath.
A number of different situations depicting the modulation of the bath
by an external noise may be physically relevant. As, for example, we consider
a simple unimolecular conversion (say, an isomerization reaction) from
$A \longrightarrow B$. The reaction is carried out in a photochemically 
active solvent under the influence of external fluctuating light intensity.
Since the fluctuations in the light intensity result in the fluctuations in
the polarization of the solvent molecules, the effective reaction field
around the reactant system gets modified. Provided the required stationarity 
of this nonequilibrium open system is maintained (which is not difficult 
in view of the experiments performed in the studies of external noise-induced 
transitions in photochemical systems \cite{rd30a}) the dynamics of barrier
crossing becomes amenable to the present theoretical analysis that follows.

The remaining part of this paper is organized as follows: In Sec.~{II} we
discuss a system-reservoir model where the later is modulated by an external
noise and establish an interesting connection between the dissipation of
the system and the response function of the reservoir due to external noise.
The stochastic motion in a linearized potential field is described in terms
of a Fokker-Planck equation in Sec.~{III}. Based on the traditional flux
over population method \cite{rd31} we derive in Sec.~{IV} the generalized
expression for the Krmaers' rate of escape from a metastable well. The 
general theory is illustrated with a specific example in Sec.~{V}. 
The paper is concluded in Sec.~{VI}.

\section{ The system-reservoir model : The reservoir modulated by external
noise 
\label{sd2}
}

We consider a classical particle of mass $M$ is coupled to a heat bath of $N$
harmonic oscillators. The various modes of heat bath are perturbed by an 
external random force. The interaction between external force and bath 
co-ordinates is added as a new term ($H_{int}$) in the standard 
system-reservoir Hamiltonian of Zwanzig form \cite{rd22}. The total 
Hamiltonian is given by
\begin{equation}
\label{eq1}
H = \frac{p^2}{2M} + V(x) + \frac{1}{2} \sum_{i=1}^N \left \{ 
\frac{p_i^2}{m_i} + m_i \omega_i^2 (q_i - a_i (x) )^2 \right \} + H_{int}
\; \; .
\end{equation}

\noindent
In Eq.(\ref{eq1}), $x$ and $p$ are the co-ordinate and momentum of the system
particle; $(q_i, p_i)$ are the variables associated with the $i$-th oscillator
and $\omega_i$ and $m_i$ are the corresponding frequency and mass 
respectively. $a_i(x)$ measures the
interaction between the particle and the bath. $V(x)$ is the potential
energy of the particle. $H_{int}$ is assumed to be of the form
\begin{equation}
\label{eq2}
H_{int} = \frac{1}{2} \sum_{i=1}^N \kappa_i \; q_i \; \epsilon (t) \;\; .
\end{equation}

\noindent
The coupling function $\kappa_i$ measures the strength of interaction and 
$\epsilon (t)$ is the external noise which we assume to be stationary and
Gaussian with zero mean, i.e., $\langle \epsilon (t) \rangle = 0$.

We now consider the interaction between the system and the heat bath to be
linear, i.e.,
\begin{equation}
\label{eq3}
a_i (x) = g_i \; x \; \; ,
\end{equation}

\noindent
and then eliminate the bath degrees of freedom in the usual way 
\cite{rd20,rd22,rd23} to obtain 
the following generalized Langevin equation
\begin{eqnarray}
\label{eq4}
\dot{x} & = & v \; \; ,\nonumber \\
\dot{v} & = & -\frac{dV}{dx} - \int_0^t dt' \; \gamma (t-t') \; v(t') + f (t)
+ \pi (t) 
\end{eqnarray}

\noindent
[ while constructing Eq.(\ref{eq4}) we have set $M$ and $m_i$ equal to
unity ] where
\begin{equation}
\label{eq5}
\gamma (t) = \sum_{i=1}^N g_i^2 \; \omega_i^2 \; \cos \omega_i t \; \; .
\end{equation}

\noindent
$f(t)$ is the fluctuation generated through the coupling between the
system and the heat bath and is given by
\begin{equation}
\label{eq6}
f (t) = \sum_{i=1}^N g_i \; \left \{ \left [ q_i (0) - g_i x(0) \right ]
\; \omega_i^2 \; \cos \omega_i t + v_i (0) \; \omega_i \; \sin \omega_i t
\right \} \; \; .
\end{equation}

\noindent
$f(t)$ thus depends on the coupling functions $g_i$ and the normal mode initial
conditions, which we assume to be canonically distributed. The statistical 
properties of $f(t)$ can then be summarized in the following equations;
\begin{mathletters}
\begin{eqnarray}
\label{eq7}
\langle f(t) \rangle & = & 0 \; \; {\rm and} \\ 
\langle f(t) f(t') \rangle & = & k_B T \gamma (t-t') \; \; .
\end{eqnarray}
\end{mathletters}

\noindent
In Eq.(\ref{eq4}), $\pi (t)$ is a force term driven by the external noise 
$\epsilon (t)$ and is given by
\begin{equation}
\label{eq8}
\pi (t) = - \int_0^t \varphi (t-t') \; \epsilon (t') \; dt' \; \; ,
\end{equation}

\noindent
where
\begin{equation}
\label{eq9}
\varphi (t) = \sum_{i=1}^N g_i \; \kappa_i \; \omega_i \sin \omega_i t \; \; .
\end{equation}

\noindent
The statistical properties of $\pi (t)$ are determined by the normal mode 
distribution of the bath, the coupling of the system with the bath, the 
coupling of the bath with the external noise and the external noise itself.
It may also be noted that the external noise $\epsilon (t)$ and the internal 
noise $ f(t)$ are independent of each other because they have been generated 
from different origin. Eq.(\ref{eq8}) is reminiscent of the familiar linear
relation between the polarization and the external field where $\pi$ and
$\epsilon$ play the role of the former and the later, respectively. 
$\varphi (t)$
can then be interpreted as a response function of the reservoir due to external
noise $\epsilon (t)$.

In the continuum limit $\gamma (t)$ and $\varphi (t)$ reduce to the following 
forms
\begin{equation}
\label{eq10}
\gamma (t)  =  \int d\omega \; {\cal D} (\omega) \; g^2 (\omega) \; 
\omega^2 \; \cos \omega t 
\end{equation}

\noindent
and
\begin{equation}
\label{eq11}
\varphi (t) =  \int d\omega \; {\cal D} (\omega) \; \kappa (\omega) \; 
\omega \; g (\omega) \; \sin \omega t \; \; .
\end{equation}

\noindent
where ${\cal D} (\omega )$ is the density of modes of the heat bath.
If the coupling functions $g(\omega )$ and $\kappa (\omega )$ are assumed to
be of the following forms \cite{rd32} to obtain a finite result in the
continuum limit,
\begin{eqnarray*}
g(\omega ) = \frac{g_0}{\sqrt{\tau_c} \; \omega} \; \; {\rm and} \; \; 
\kappa (\omega ) = \sqrt{\tau_c} \; \omega \; \kappa_0 
\end{eqnarray*}

\noindent
where $g_0$ and $\kappa_0$ are constants and $\tau_c$ is the correlation 
time, i.e., $\tau_c^{-1}$ is the cutoff frequency of the harmonic 
oscillators then the expressions for $\gamma (t)$ and $\varphi (t)$ reduce
to
\begin{equation}
\label{eq12}
\gamma (t)  =  \frac{g_0^2}{\tau_c} \;
\int d\omega \; {\cal D} (\omega) \; \cos \omega t 
\end{equation}

\noindent
and
\begin{equation}
\label{eq13}
\varphi (t)  =  g_0 \; \kappa_0 \; \int d\omega \; {\cal D} (\omega) \; 
\omega \; \sin \omega t \; \; .
\end{equation}

\noindent
From the above two relations, we obtain
\begin{equation}
\label{eq14}
\frac{d\gamma}{dt} = -\frac{g_0}{\kappa_0} \; \frac{1}{\tau_c} \varphi (t) \; \; .
\end{equation}

\noindent
Eq.(\ref{eq14}) is an important content of the present model. This expresses
how the dissipative kernel $\gamma (t)$ depends on the response function
$\varphi (t)$ of the medium due to external noise $\epsilon (t)$
[ see Eq.(\ref{eq8}) ]. Such a relation for the open system can be anticipated
in view of the fact that both the dissipation and the response function 
crucially depend on the properties of the reservoir especially on its density
of modes and its coupling to the system and the external noise source.
In what follows we shall be concerned with the 
consequences of this relation in terms of the Langevin description in the 
next section ( Eq.(\ref{eq15}) ). 

\section{ Generalized Fokker-Planck description of the linearized motion 
\label{sd3}
}

We now consider the system to be a harmonically bound particle of unit mass
and of frequency $\omega_0$. Then because of Eq.(\ref{eq8}) the Langevin 
equation (\ref{eq4}) becomes
\begin{eqnarray}
\label{eq15}
\dot{x} & = & v \; \; , \nonumber \\
\dot{v} & = & -\omega_0^2 x - \int_0^t dt' \; \gamma (t-t') \; v(t') + f (t)
- \int_0^t dt' \; \varphi (t-t') \; \epsilon (t')
\end{eqnarray}

\noindent
The Laplace transform of Eq. (\ref{eq15}) allows us to write a formal 
solution for the displacement of the form
\begin{eqnarray}
\label{eq16}
x (t) & = & \langle x (t) \rangle + \int_0^t dt' \; h (t-t')\; f (t') -
\frac{\kappa_0}{g_0}\tau_c \omega_0^2 \int_0^t dt' \; h (t-t')\; \epsilon (t')
\nonumber \\
& & - \frac{\kappa_0}{g_0}\tau_c \int_0^t dt' \; h_2 (t-t')\; \epsilon (t')
\; \; ,
\end{eqnarray}

\noindent
where we have made use of the relation (\ref{eq14}) explicitly.

\noindent
Here
\begin{equation}
\label{eq17}
\langle x (t) \rangle = \chi_x (t) x (0) + h (t) v (0)
\end{equation}

\noindent
with $x(0)$ and $v(0)$ being the initial position and initial velocity of the 
oscillator, respectively, which are nonrandom and
\begin{equation}
\label{eq18}
\chi_x (t) = \left [ 1 - \omega_0^2 \int_0^t h (\tau) \; d\tau \right ] 
\; \; .
\end{equation}

\noindent
The kernel $h(t)$ is the Laplace inversion of 
\begin{equation}
\label{eq19}
\tilde{h} (s) = \frac{1}{s^2 + \tilde{\gamma} (s) \; s + \omega_0^2}
\end{equation}

\noindent
where, $\tilde{\gamma} (s) = \int_0^\infty e^{-st} \; \gamma (t) \; dt$, is 
the Laplace transform of the friction kernel $\gamma (t)$, and
\begin{equation}
\label{eq20}
h_2 (t) = \frac{d^2 h(t)}{dt^2} \; \;.
\end{equation}

\noindent
The time derivative of Eq.(\ref{eq16}) yields
\begin{eqnarray}
\label{eq21}
v (t) & = & \langle v (t) \rangle + \int_0^t dt' \; h_1 (t-t')\; f (t') -
\frac{\kappa_0}{g_0}\tau_c \omega_0^2 \int_0^t dt' \; h_1 (t-t')\; \epsilon (t')
\nonumber \\
& & - \frac{\kappa_0}{g_0}\tau_c \int_0^t dt' \; h_3 (t-t')\; \epsilon (t')
\end{eqnarray}

\noindent
where
\begin{equation}
\label{eq22}
\langle v (t) \rangle = -\omega_0^2 h(t) + v (0) h_1 (t) \; \; ,
\end{equation}

\begin{equation}
\label{eq23}
h_1 (t) = \frac{d h(t)}{dt}
\end{equation}

\noindent
and
\begin{equation}
\label{eq24}
h_3 (t) = \frac{d^3 h(t)}{dt^3} \; \;.
\end{equation}

Next we calculate the variances. From the formal solution of $x(t)$ and
$v(t)$, the explicit expressions for the variances are obtained which are
given below;
\begin{eqnarray}
\label{eq25}
\sigma_{xx}^2 (t) & = & \langle [ x (t) - \langle x (t) \rangle ]^2 \rangle
\nonumber \\
& = & 2 \int_0^t dt_1 \; h (t_1) \int_0^{t_1} dt_2 \; h (t_2) \; 
\langle f(t_1) f(t_2) \rangle \nonumber \\
& & + 2 \left ( \frac{\kappa_0}{g_0} \tau_c \omega_0^2 \right )^2
\int_0^t dt_1 \; h (t_1) \int_0^{t_1} dt_2 \; h (t_2) \; 
\langle \epsilon (t_1) \epsilon (t_2) \rangle \nonumber \\
& & + 2 \left ( \frac{\kappa_0}{g_0} \tau_c \right )^2
\int_0^t dt_1 \; h_2 (t_1) \int_0^{t_1} dt_2 \; h_2 (t_2) \; 
\langle \epsilon (t_1) \epsilon (t_2) \rangle \nonumber \\
& & + 2 \left ( \frac{\kappa_0}{g_0} \tau_c \right )^2 \omega_0^2 \;
\int_0^t dt_1 \; h (t_1) \int_0^{t_1} dt_2 \; h_2 (t_2) \; 
\langle \epsilon (t_1) \epsilon (t_2) \rangle \; \; ,
\end{eqnarray}

\begin{eqnarray}
\label{eq26}
\sigma_{vv}^2 (t) & = & \langle [ v (t) - \langle v (t) \rangle ]^2 \rangle
\nonumber \\
& = & 2 \int_0^t dt_1 \; h_1 (t_1) \int_0^{t_1} dt_2 \; h_1 (t_2) \; 
\langle f(t_1) f(t_2) \rangle \nonumber \\
& & + 2 \left ( \frac{\kappa_0}{g_0} \tau_c \omega_0^2 \right )^2
\int_0^t dt_1 \; h_1 (t_1) \int_0^{t_1} dt_2 \; h_1 (t_2) \; 
\langle \epsilon (t_1) \epsilon (t_2) \rangle \nonumber \\
& & + 2 \left ( \frac{\kappa_0}{g_0} \tau_c \right )^2
\int_0^t dt_1 \; h_3 (t_1) \int_0^{t_1} dt_2 \; h_3 (t_2) \; 
\langle \epsilon (t_1) \epsilon (t_2) \rangle \nonumber \\
& & + 2 \left ( \frac{\kappa_0}{g_0} \tau_c \right )^2 \omega_0^2 \;
\int_0^t dt_1 \; h_1 (t_1) \int_0^{t_1} dt_2 \; h_3 (t_2) \; 
\langle \epsilon (t_1) \epsilon (t_2) \rangle \; \; ,
\end{eqnarray}

\noindent
and
\begin{eqnarray}
\label{eq27}
\sigma_{xv}^2 (t)  & = & 
\langle [ x (t) - \langle x (t) \rangle ]
[ v (t) - \langle v (t) \rangle ] \rangle \nonumber \\
& = & \frac{1}{2} \dot{\sigma}_{xx}^2 (t) 
\end{eqnarray}

\noindent
where we have assumed that the noises $f(t)$ and $\epsilon (t)$ are symmetric 
with respect to the time argument and have made use the fact that $f(t)$
and $\epsilon (t)$ are uncorrelated.

Due to the Gaussian property of the noises $f(t)$ and $\epsilon (t)$ and the 
linearity of the Langevin equation (\ref{eq15}), we see that the joint
probability density $p(x,v,t)$ of the oscillator must be Gaussian. The joint
characteristic function associated with the density is
\begin{equation}
\label{eq28}
\tilde{p} (\mu,\rho, t)  =  \exp \left \{ i \langle x(t) \rangle \mu +
i \langle v(t) \rangle \rho -\frac{1}{2} \left [ \sigma_{xx}^2 (t) \mu^2 +
2 \sigma_{xv}^2 (t) \rho \mu + \sigma_{vv}^2 (t) \rho^2 \right ] \right \}
\; \; .
\end{equation}

\noindent
Using the method of characteristic function \cite{rd33,rd34}
and the above expression 
(\ref{eq28}) we find the general Fokker-Planck equation associated with the 
probability density function $p(x,v,t)$ for the process (\ref{eq15});
\begin{equation}
\label{eq29}
\frac{\partial p}{\partial t} = -v \frac{\partial p}{\partial x} +
\bar{\omega}_0^2 (t) x \frac{\partial p}{\partial v} + \bar{\gamma} (t)
\frac{\partial}{\partial v} (vp) + \phi (t) \frac{\partial^2 p}{\partial v^2}
+ \psi (t) \frac{\partial^2 p}{\partial v \partial x} \; \; ,
\end{equation}

\noindent
where
\begin{eqnarray*}
\bar{\gamma} (t) & = & -\frac{d}{dt} \ln \Upsilon (t) \; \; ,\\
\bar{\omega}_0^2 (t) & = & \frac{-h(t) \; h_1 (t) + h_1^2 (t)}{\Upsilon (t)}
\; \; {\rm and} \\
\Upsilon (t) & = & \frac{h_1 (t)}{\omega_0^2} \left [ 1 - \omega_0^2
\int_0^\tau d\tau \; h (\tau ) \right ] + h^2 (t) \; \; .
\end{eqnarray*}

\noindent
The functions $\phi (t)$ and $\psi (t)$ are defined by
\begin{eqnarray}
\label{eq30}
\phi (t) & = & \bar{\omega}_0^2 (t) \sigma_{xv}^2 + \bar{\gamma} 
\sigma_{vv}^2 + \frac{1}{2} \dot{\sigma}_{xv}^2 \; \; , \nonumber \\
\psi (t) & = & \dot{\sigma}_{xv}^2 + \bar{\gamma} (t) \sigma_{xv}^2 +
\bar{\omega}_0^2 \sigma_{xx}^2 - \sigma_{vv}^2
\end{eqnarray}

\noindent
where the covariances are to be calculated for a particular given noise 
process.

For the internal noise processes Carmeli and Nitzan \cite{rd35}
and H\"anggi \cite{rd36} have shown 
that for several models the various time dependent parameters 
$\bar{\omega}_0^2 (t)$, $\bar{\gamma} (t)$, etc. do exist asymptotically as
$t \rightarrow \infty$. The above consideration shows that 
$h(t)$, $h_1(t)$, etc. do not depend on the nature of the noise but
depend only on the relaxation $\bar{\gamma} (t)$. 

We now discuss the asymptotic properties of $\phi (t)$ and $\psi (t)$,
which in turn are dependent on the variances $\sigma_{xx}^2 (t)$ and
$\sigma_{vv}^2 (t)$, as $t \rightarrow \infty$ since they play a significant
role in our further analysis that follows.

\noindent
From Eqs. (\ref{eq25}) and (\ref{eq26}), we may write
\begin{eqnarray*}
\sigma_{xx}^2 (t) = \sigma_{xx}^{2(i)} (t) + \sigma_{xx}^{2(e)} (t)
\end{eqnarray*}

\noindent
and
\begin{eqnarray*}
\sigma_{vv}^2 (t) = \sigma_{vv}^{2(i)} (t) + \sigma_{vv}^{2(e)} (t) \; \; .
\end{eqnarray*}

\noindent
where `$i$' denotes the part corresponding to internal noise $f(t)$ and
`$e$' corresponds to the external noise $\epsilon (t)$. Since the average
velocity of the oscillator is zero as $t \rightarrow \infty$ we see from
Eq.(\ref{eq22}) that $h (t)$ and $h_1 (t)$ must be zero as 
$t \rightarrow \infty$. Also from Eq.(\ref{eq17}) we observe that the function
$\chi_x (t)$ must decay to zero for long times. Hence, from Eq.(\ref{eq18})
we see that the stationary value of the integral of $h(t)$ is $1/\omega_0^2$,
i.e., 
\begin{equation}
\label{eq30a}
\int_0^\infty h (t) \; dt = \frac{1}{\omega_0^2} \; \; .
\end{equation}

\noindent
Now, $\sigma_{xx}^{2(i)} (t)$ and $\sigma_{vv}^{2(i)} (t)$ of Eqs.(\ref{eq25})
and (\ref{eq26}) can be written in the form
\begin{eqnarray}
\sigma_{xx}^{2(i)} (t) & = & 2 \int_0^t dt_1 \; h (t_1) \int_0^{t_1} dt_2 \;
h (t_2) \; \langle f (t_1) f (t_2) \rangle \nonumber \\
& = & k_BT \left [ 2 \int_0^t d\tau \; h (\tau ) - h^2 (t) -
\omega_0^2 \left \{ \int_0^t d\tau \; h (\tau ) \right \}^2 \right ]
\label{eq30b}
\end{eqnarray}

\noindent
and
\begin{eqnarray}
\sigma_{vv}^{2(i)} (t) & = & 2 \int_0^t dt_1 \; h_1 (t_1) \int_0^{t_1} dt_2 \;
h_1 (t_2) \; \langle f (t_1) f (t_2) \rangle \nonumber \\
& = & k_BT \left [ 1 - h_1^2 (t) - \omega_0^2 h^2 (t) \right ] \; \; .
\label{eq30c}
\end{eqnarray}

\noindent
From the above two expressions [ Eqs.(\ref{eq30b}) and (\ref{eq30c}) ]
we see that 
\begin{equation}
\label{eq30d}
\sigma_{xx}^{2(i)} (\infty) = \frac{k_BT}{\omega_0^2} \; \; \;
{\rm and} \; \; \; \sigma_{vv}^{2(i)} (\infty) = k_BT \; \; .
\end{equation}

\noindent
It is important to note that these stationary values are not related to the 
intensity and correlation time of the internal noise.

We next consider the parts,  $\sigma_{xx}^{2(e)} (t)$ and 
$\sigma_{vv}^{2(e)} (t)$, due to the presence of the external noise. The
Laplace transform of Eq.(\ref{eq16}) yields the expression
\begin{equation}
\label{eq30e}
\tilde{x} (s) - \langle \tilde{x} (s) \rangle = \tilde{h} (s) \tilde{f} (s)
- \frac{\kappa_0}{g_0} \tau_c \omega_0^2 \tilde{h} (s) \tilde{\epsilon} (s)
- \frac{\kappa_0}{g_0} \tau_c s^2 \tilde{h} (s) \tilde{\epsilon} (s)
\end{equation}

\noindent
where
\begin{eqnarray}
\langle \tilde{x} (s) \rangle & = & \left \{ \frac{1}{s} - 
\frac{\omega_0^2}{s [ s^2 + s \tilde{\gamma} (s) + \omega_0^2] } \right \}
x (0) + \frac{1}{s^2 + s \tilde{\gamma} (s) + \omega_0^2 } \; v (0)
\nonumber \\
& = & \left \{ \frac{1}{s} - \omega_0^2 \frac{ \tilde{h} (s)}{s} \right \}
x (0) + \tilde{h} (s) v (0) \; \; .
\label{eq30f}
\end{eqnarray}

\noindent
From the above equation (\ref{eq30e}) we can calculate the variance
$\sigma_{xx}^2$ in the Laplace-transformed  space which can be
identified as the Laplace 
transform of Eq.(\ref{eq25}). Thus, for the part $\sigma_{xx}^{2(e)} (t)$
we observe that,
$\tilde{\sigma}_{xx}^{2 (e)} (s)$ contains terms like
$\left ( \frac{\kappa_0}{g_0} \tau_c \omega_0^2 \tilde{h} (s) \right )^2 
\langle \tilde{\epsilon}^2 (s) \rangle$.  
Since, we have assumed the
stationarity of the noise $\epsilon (t)$, we conclude that if 
$\tilde{C} (0)$ exists [ where 
$C (t-t') = \langle \epsilon (t) \epsilon (t') \rangle $ ],  then the stationary
value of $\sigma_{xx}^{2(e)} (t)$ exists and becomes a constant that
depends on the correlation time and the strength of the noise. Similar
argument is also valid for $\sigma_{vv}^{2(e)} (t)$.

Summarizing the above discussions we note that,

\noindent
(i) the internal noise-driven parts of $\sigma_{xx}^2 (t)$ and 
$\sigma_{vv}^2 (t)$, i.e., $\sigma_{xx}^{2(i)} $ and $\sigma_{vv}^{2(i)}$
approach the fixed values which are independent of the noise correlation and 
the intensity as $t \rightarrow \infty$, 

\noindent
(ii) the external noise driven parts of variances also approach the
constant values at the stationary ($t \rightarrow \infty$) limit which are
dependent on the strength and the correlation time of the noise.

Hence we conclude, following the Refs.[36,37] and our preceding discussions
that even in presence of an
external noise the coefficients of the Fokker-Planck equation (\ref{eq29}) 
do exist asymptotically and we write its
steady state version for the asymptotic values of the parameters as,
\begin{equation}
\label{eq31}
-v \frac{\partial p}{\partial x} +
\bar{\omega}_0^2 x \frac{\partial p}{\partial v} + \bar{\gamma}
\frac{\partial}{\partial v} (vp) + \phi (\infty) \frac{\partial^2 p}{\partial v^2}
+ \psi (\infty) \frac{\partial^2 p}{\partial v \partial x} = 0 \; \; ,
\end{equation}

\noindent
where, $\bar{\omega}_0^2$, $\bar{\gamma}$, $\phi (\infty )$, $\psi (\infty )$,
etc. are to be calculated from the general definition (\ref{eq30}) for the
steady state.

The general steady state solution of the above equation (\ref{eq31}) is
\begin{equation}
\label{eq32}
p_{st} (x,v) = \frac{1}{Z} \exp \left [ - \left \{ \frac{v^2}{2D_0} +
\frac{\bar{\omega}_0^2 x}{ 2 ( D_0 + \psi (\infty) \; )} \right \} \right ]
\end{equation}

\noindent
where
\begin{equation}
\label{eq33}
D_0 = \frac{ \phi (\infty) }{ \bar{\gamma} }
\end{equation}

\noindent
and $Z$ is the normalization constant. The solution (\ref{eq32}) can be
verified by direct substitution. The distribution (\ref{eq32}) is not an 
equilibrium distribution. This stationary distribution for the 
nonequilibrium open system 
plays the role of an equilibrium distribution of the closed system which
may, however, be recovered in the absence of external noise term. 

Some further pertinent points regarding rate theory for nonequilibrium
systems may be in order. It is well-known that the equilibrium state of a 
closed thermodynamic system with homogenous boundary conditions is 
time-independent. The open, e.g., the driven system on the other hand, may 
show up the complicated spatio-temporal structures or may settle down to
multiple steady states \cite{rd30a}. The external noise may then induce 
transitions between them. It is, however, important to realize that these 
features originate only when one takes into account of the nonlinearity of 
the system
in full and the external noise drives the system directly. Secondly, in
most open nonequilibrium systems the lack of detail balance symmetry gives
rise to severe problem in the determination of the stationary probability
distribution for multidimensional problems \cite{rd37}. At this juncture three
points are to be noted. First, for the present problem we have made use of the 
linearization of the potential at the bottom and at top of the barrier
(as is done in Kramers' \cite{rd1} and in most of the post-Kramers' development
\cite{rd7,rd8,rd9,rd16,rd35}) which precludes the existence of a multiple steady states. 
Second, the external noise considered here drives the bath rather than the
system directly. Third, the problem is one-dimensional. Thus, an unique
stationary probability density which is an essential requirement for the
mean first passage time or flux over population method (as in the present 
case) for the calculation of rate and which is readily obtainable in the
case of closed equilibrium system, can also be obtained for this open
nonequilibrium system.

\section{ Kramers' escape rate
\label{sd4}
}

We now turn to the problem of decay of a metastable state. To this end we
consider as usual a `Brownian particle' moving in a one-dimensional double
well potential $V (x)$ \cite{rd1,rd16}. In Kramers approach \cite{rd1}, the
particle coordinate $x$ corresponds to the reaction coordinate and its values
at the minima of the potential $V(x)$ denotes the reactant and product
states. The maxima of $V(x)$ at $x_b$ separating these states 
corresponds to the activated complex. All the remaining
degrees of freedom of both reactant and solvent molecules constitute a heat
bath at temperature $T$. Our object is to
calculate the essential modification of Kramers' rate 
when the bath modes are perturbed by an external random force
when the system has attained a steady state.

Linearizing the motion around barrier top at $x=x_b$ the Langevin equation 
(\ref{eq4}) can be written down as
\begin{eqnarray}
\label{eq34}
\dot{y} & = & v \; \; ,\nonumber \\
\dot{v} & = & \omega_b^2 \; y - \int_0^t dt' \; \gamma (t-t') \; v(t') 
+ f (t) + \pi (t) \; \; ,
\end{eqnarray}

\noindent
where, $y=x-x_b$ and the barrier frequency $\omega_b^2$ is defined by
\begin{equation}
\label{eq35}
V (y) = V_b - \frac{1}{2} \omega_b^2 y^2 \; \; ; \; \; \omega_b^2 > 0
\; \; .
\end{equation}

\noindent
Correspondingly the motion of the particle is governed by the 
Fokker-Planck equation (\ref{eq29})
\begin{equation}
\label{eq36}
\frac{\partial p}{\partial t} = -v \frac{\partial p}{\partial y} -
\bar{\omega}_b^2 (t) y \frac{\partial p}{\partial v} + \bar{\gamma}_b (t)
\frac{\partial}{\partial v} (vp) + \phi_b (t) \frac{\partial^2 p}{\partial v^2}
+ \psi_b (t) \frac{\partial^2 p}{\partial v \partial y} \; \; ,
\end{equation}

\noindent
where, the suffix `$b$' indicates that all the coefficients are to be 
calculated using the general definition (\ref{eq30}) for the barrier top 
region.

It is apparent from Eqs.(\ref{eq31}) and (\ref{eq36}) that since the dynamics
is non-Markovian and the system is thermodynamically open one has to deal
with the renormalized frequencies $\bar{\omega}_0$ and $\bar{\omega}_b$ near
the bottom or top of the well, respectively. Following Kramers \cite{rd1} we
make the ansatz that the nonequilibrium, steady state probability 
$p_b $, generating a nonvanishing diffusion current $j$, across the 
barrier is given by
\begin{equation}
\label{eq37}
p_b (x,v) = \exp \left [ - \left \{ \frac{v^2}{2D_b} +
\frac{ \tilde{V} (x) }{ D_b + \psi_b (\infty) } \right \} \right ]
\; \xi (x,v)
\end{equation}

\noindent
where
\begin{equation}
\label{eq38}
D_b = \frac{ \phi_b (\infty ) }{\bar{\gamma}_b} \; \; .
\end{equation}

\noindent
$\tilde{V} (x)$ is the renormalized linear potential as
\begin{eqnarray}
\label{eq39}
\tilde{V} (x) & = & V (x_0) + \frac{1}{2}\bar{\omega}_0^2 (x-x_0)^2 \; \; ,
\; \; {\rm near \; the \; bottom} \nonumber \\
\tilde{V} (x) & = & V (x_b) - \frac{1}{2}\bar{\omega}_b^2 (x-x_b)^2 \; \; ,
\; \; {\rm near \; the \; top} 
\end{eqnarray}

\noindent
with $\bar{\omega}_0^2$, $\bar{\omega}_b^2 > 0$. The unknown function 
$\xi (x,v)$ obeys the natural boundary condition that for 
$x \rightarrow \infty$, $\xi (x,v)$ vanishes.

The ansatz of the form (\ref{eq37}) denoting the steady state distribution
is motivated by the local analysis near the bottom and the top of the
barrier in the Kramers' sense \cite{rd1}. For a stationary nonequilibrium
system, on the other hand, the relative population of the two regions, in 
general, depends on the global properties of the potential leading to an
additional factor in the rate expression. Although because of the Kramers'
type ansatz \cite{rd1}which is valid for the local analysis, such a 
consideration is outside the scope of the present treatment, we point out 
a distinctive feature in the ansatz (\ref{eq37}) compared to Kramers' ansatz.
While in the latter case one considers a complete factorization of the 
equilibrium
part (Boltzmann) and the dynamical part, the ansatz (\ref{eq37}) incorporates
the additional dynamical contribution through dissipation and strength of
the noise into the exponential part. This modification of Kramers' ansatz
(by dynamics) is due to nonequilibrium nature of the system.
Thus unlike Kramers', the exponential
factors in (\ref{eq37}) and in the stationary distribution (\ref{eq32})
which serves as a boundary condition are markedly different. While a global 
analysis may even modify the standard Kramers' result our aim here is to
understand the modification of the rate due to modulation of the bath
driven by an external noise, within the perview of Kramers' type ansatz. The
internal consistency of the treatment, however, can be checked by recovering
the Kramers' result when the external noise is switched off.

From equation (\ref{eq36}), using (\ref{eq37}) we obtain the equation for 
$\xi (y,v)$ in the steady state in the neighborhood of $x_b$, the equation
\begin{equation}
\label{eq40}
-\left ( 1+ \frac{\psi_b (\infty) }{D_b} \right ) v 
\frac{\partial \xi}{\partial y}
- \left [ \frac{D_b}{D_b+\psi_b (\infty)} 
\bar{\omega}_b^2 y + \bar{\gamma}_b v \right ] 
\frac{\partial \xi}{\partial v} + 
\phi_b (\infty ) \frac{\partial^2 \xi}{\partial v^2} +
\psi_b (\infty ) \frac{\partial^2 \xi}{\partial v \partial y} = 0 \; \; .
\end{equation}

\noindent
We then make use of the following transformation
\begin{eqnarray*}
u = v + a \; y \; \; \; , \; \; y = x - x_b
\end{eqnarray*}

\noindent
where $a$ is a constant to be determined. We obtain from Eq.(\ref{eq40})
\begin{equation}
\label{eq41}
\{ \phi_b (\infty) + a \psi_b (\infty) \; \} \frac{d^2 \xi}{d u^2}
- \left [ \frac{D_b}{D_b+\psi_b (\infty) } \bar{\omega}_b^2 x + \left \{
\bar{\gamma}_b + a \left ( 1 + \frac{\psi_b (\infty) }{D_b} \right ) 
\right \} v \right ] \frac{d \xi}{d u} = 0 \; \; .
\end{equation}

\noindent
Putting
\begin{equation}
\label{eq42}
\frac{D_b}{D_b+\psi_b (\infty )} \bar{\omega}_b^2 x + \left \{ \bar{\gamma}_b 
+ a \left ( 1 + \frac{\psi_b (\infty) }{D_b} \right ) \right \} v 
= -\lambda u \; \; ,
\end{equation}

\noindent
(where $\lambda$ being another constant to be determined) we obtain the 
ordinary differential equation for $\xi (u)$
\begin{equation}
\label{eq43}
\frac{d^2 \xi}{d u^2} + \Lambda u 
\frac{d \xi}{d u} = 0
\end{equation}

\noindent
where
\begin{equation}
\label{eq44}
\Lambda = \frac{\lambda}{\phi_b (\infty) + a \psi_b (\infty) } \; \; .
\end{equation}

\noindent
and the two constants $\lambda$ and $a$  must satisfy the simultaneous
relations
\begin{mathletters}
\begin{eqnarray}
-\lambda a & = & \frac{D_b}{D_b + \psi_b (\infty) } \bar{\omega}_b^2 \; \; ,
\\
-\lambda & = & \bar{\gamma}_b + a \left ( 1 + \frac{\psi_b (\infty) }{D_b}
\right ) \; \; .
\end{eqnarray}
\end{mathletters}

\noindent
This implies that, the constant $a$ must satisfy the quadratic equation
\begin{equation}
\label{eq45}
\frac{D_b + \psi_b (\infty) }{D_b} \; a^2 + \bar{\gamma}_b \; a -
\frac{D_b}{D_b + \psi_b (\infty) } \; \bar{\omega}_b^2 = 0 
\end{equation}

\noindent
which allows
\begin{equation}
\label{eq46}
a_\pm = \frac{D_b}{2 ( D_b + \psi_b (\infty) \; ) } \left \{ -\bar{\gamma}_b 
\pm \sqrt{ \bar{\gamma}_b^2 + 4 \bar{\omega}_b^2 } \right \} \; \; .
\end{equation}

\noindent
The general solution of Eq.(\ref{eq43}) is
\begin{equation}
\label{eq47}
\xi (u) = F_2 \int_0^u \exp \left ( - \frac{\Lambda \; z^2}{2} \right ) \;
dz + F_1
\end{equation}

\noindent
where $F_1$ and $F_2$ are constants of integration. We look for a solution
which vanishes for large $x$. For this to happen the integral in 
(\ref{eq47}) should remain finite for $|u| \rightarrow +\infty$. This implies
that $\Lambda > 0$ so that only $a_-$ becomes
relevant. Then the requirement $p_b (x,v) \rightarrow 0$ for 
$x \rightarrow +\infty$ yields
\begin{equation}
\label{eq48}
F_1 = F_2 \sqrt{ \frac{\pi}{2\Lambda} } \; \; .
\end{equation}

\noindent
Thus we have
\begin{eqnarray*}
\xi (u) = F_2 \left [ \sqrt{ \frac{\pi}{2\Lambda} } + 
\int_0^u \exp \left ( - \frac{\Lambda \; z^2}{2} \right ) \;
dz  \right ]
\end{eqnarray*}

\noindent
and correspondingly
\begin{equation}
\label{eq49}
p_b (x,v) =  F_2 \left [ \sqrt{ \frac{\pi}{2\Lambda} } + 
\int_0^u \exp \left ( - \frac{\Lambda \; z^2}{2} \right ) \;
dz  \right ] \; 
\exp \left [ - \left \{ \frac{v^2}{2D_b} +
\frac{ \tilde{V} (x) }{ D_b + \psi_b (\infty) } \right \} \right ] \; \; .
\end{equation}

The current across the barrier associated with this distribution is given by
\begin{eqnarray*}
j = \int_{-\infty}^{+\infty} v \; p_b (x=x_b, v) \; dv
\end{eqnarray*}

\noindent
which may be evaluated using (\ref{eq49}) and the linearized version of
$\tilde{V} (x)$, namely, 
$\tilde{V} (x)  =  V (x_b) - \frac{1}{2}\bar{\omega}_b^2 (x-x_b)^2 $, as
\begin{equation}
\label{eq50}
j = F_2 \; \left ( \frac{2\pi}{\Lambda + D_b^{-1} } \right )^{1/2} \; D_b
\; \exp \left [ -\frac{ V (x_b) }{ D_b + \psi_b (\infty) } \right ] \; \; .
\end{equation}

\noindent
To determine the remaining constant $F_2$ we proceed as follows. We first 
note that as $x\rightarrow - \infty$ the pre-exponential factor in 
Eq.(\ref{eq49}) reduces to the following form
\begin{equation}
F_2 [ .... ] = F_2 \; \left ( \frac{2 \pi}{\Lambda} \right )^{1/2}
\; \; .
\end{equation}

\noindent
We then obtain the reduced distribution function in $x$ as
\begin{equation}
\label{eq51}
\tilde{p}_b (x\rightarrow -\infty) = 2\pi F_2 \; 
\left ( \frac{D_b}{\Lambda} \right )^{1/2} \;
\exp \left [ -\frac{ \tilde{V} (x) }{ D_b + \psi_b (\infty) } \right ] 
\; \; ,
\end{equation}

\noindent
where we have used the definition for the reduced distribution as
\begin{eqnarray*}
\tilde{p} (x) = \int_{-\infty}^{+\infty} p (x,v) \; dv\; \; .
\end{eqnarray*}

\noindent
Similarly we derive the reduced distribution in the left well around
$x \approx x_0$ using Eq.(\ref{eq32}) where the linearized potential is
$\tilde{V} (x)  =  V (x_0) + \frac{1}{2}\bar{\omega}_0^2 (x-x_0)^2 $,
\begin{equation}
\label{eq52}
\tilde{p}_{st} (x) = \frac{1}{Z} \; \sqrt{2\pi D_0} \;
e^{ -\frac{V(x_0)}{D_0 + \psi (\infty)} } \; 
\exp \left [ - \frac{ \bar{\omega}_0^2 (x-x_0)^2 }{ 2(D_0 + \psi (\infty) \;)} 
\right ]  \; \; ,
\end{equation}

\noindent
with the normalization constant $1/Z$
\begin{eqnarray*}
\frac{1}{Z} = \frac{\bar{\omega}_0}{2\pi \sqrt{D_0 (D_0 +\psi (\infty) ) } }
\; e^{ \frac{V(x_0)}{D_0 + \psi (\infty)} } \; \; .
\end{eqnarray*}

The comparison of the distributions (\ref{eq51}) and (\ref{eq52}) near
$x=x_0$, i.e., 
\begin{equation}
\tilde{p}_{st} (x_0) = \tilde{p}_b (x_0)
\end{equation}

\noindent
gives
\begin{equation}
\label{eq53}
F_2 = \left ( \frac{\Lambda}{D_b} \right )^{1/2} \; 
\frac{\bar{\omega}_0}{2\pi \sqrt{ 2\pi ( D_0 +\psi (\infty) ) } }
\; \exp \left [ \frac{ V(x_0) }{ D_b + \psi_b (\infty) } \right ]  \; \; .
\end{equation}

\noindent
Hence from (\ref{eq50}), the normalized current or the barrier crossing rate 
$k$ is given by
\begin{equation}
\label{eq54}
k = \frac{ \bar{\omega}_0 }{2\pi } \; 
\frac{ D_b }{ \{ D_0 + \psi (\infty) \}^{1/2} } \;
\left ( \frac{\Lambda}{1+ \Lambda D_b} \right )^{1/2} \; 
\exp \left [ -\frac{ E_0 }{ D_b + \psi_b (\infty) } \right ]
\end{equation}

\noindent
where $E_0$ is the activation energy, $E_0 = V(x_b) - V(x_0)$. Since the
temperature due to internal thermal noise, the strength of the external 
noise and the damping constant are buried in the parameters $D_0$, $D_b$,
$\psi_0$, $\psi_0$ and $\Lambda$ the generalized expression look somewhat 
cumbersome. We point out that the subscripts `$0$' and `$b$' in $D$ and
$\psi$ refer to the well or barrier top region, respectively. Eq.(\ref{eq54})
is the central result of this chapter. The dependence 
of the rate on the parameters can be exposed explicitly once we consider
the limiting cases.

\section{Example : $\delta$-correlated external noise and Ornstein-Uhlenbeck
internal noise 
\label{sd5}
}

We consider a particular case where the external noise $\epsilon(t)$ is  
$\delta$-correlated and the internal noise is an Ornstein-Uhlenbeck (O-U) 
process, i.e.,
\begin{eqnarray*}
\langle \epsilon (t) \epsilon (t') \rangle = 2 D \delta (t-t')
\end{eqnarray*}

\noindent
and
\begin{eqnarray*}
\langle f(t) f(t') \rangle = g_0^2 k_BT \frac{ e^{-|t-t'|/\tau_c} }{\tau_c}
\; \; .
\end{eqnarray*}

\noindent
One can trace the origin of the above correlation function for internal noise
by considering a Lorentzian type frequency distribution of the normal mode 
variables and Eq.(\ref{eq6}). Consequently,
from the fluctuation-dissipation relation we derive the dissipative kernel as
\begin{equation}
\label{eq55}
\gamma (t-t') = g_0^2 \frac{ e^{-|t-t'|/\tau_c} }{\tau_c} \; \; .
\end{equation}

\noindent
It should be noted that for $\tau_c \rightarrow 0$, the above noise processes
become $\delta$-correlated.

The correlation functions as given above and the dissipative kernel provide
the required quantities for the calculation of $p_{st} (x,v)$ 
[ Eq.(\ref{eq32}) ] and the generalized rate expression (\ref{eq54}). Thus
we have [see the Appendix]
\begin{equation}
\phi (\infty ) = \bar{\gamma} ( k_BT + D \kappa_0^2 ) \; \; 
{\rm and} \; \; \psi (\infty ) = 0 \; \; .
\end{equation}

\noindent
which gives
\begin{equation}
D_0 = \frac{ \phi (\infty ) }{ \bar{\gamma} } = k_BT + D \kappa_0^2
\; \; .
\end{equation}

\noindent
Hence from equation (\ref{eq32}) we see that the steady state distribution
is given by
\begin{equation}
\label{eq56}
p_{st} (x,v) = \frac{1}{Z} \exp \left [ -
\frac{ \omega_0^2 x^2 + v^2 }{ 2 ( k_BT + D \kappa_0^2 ) } \right ] \; \; ,
\end{equation}

\noindent
since for a Markovian process  $\bar{\omega}_0^2 = \omega_0^2$. From 
Eq.(\ref{eq56}) we see that, the steady state probability density in our 
example does not depend on the correlation time of noise but does depend on
the strength and coupling of the external noise. The result is in agreement
with the one obtained earlier by Bravo et. al. \cite{rd32}

We now return to our generalized rate expression Eq.(\ref{eq54}).
For the present case we have computed [see the Appendix]
\begin{mathletters}
\begin{eqnarray}
\label{eq57}
\psi (\infty ) & = & \psi_b ( \infty ) = 0  \; \; , \\
D_0 & = & D_b = k_BT + D \kappa_0^2 \; \; , \\
\bar{\omega}_0^2 & = & \omega_0^2 \; , \; \bar{\omega}_b^2  = \omega_b^2
\; \; , \\
\Lambda & = & \frac{\lambda}{g_0^2 ( k_BT + D \kappa_0^2 )} \; \; {\rm and} 
\\
a_- & = & -\frac{g_0^2}{2} - \sqrt{ \frac{g_0^4}{4} + \omega_b^2 } \; \; .
\end{eqnarray}
\end{mathletters}

\noindent
Using all these values, we obtain from Eq.(\ref{eq54})
\begin{equation}
\label{eq58}
k = \frac{\omega_0}{2\pi \omega_b} \left [ \left \{ \frac{g_0^4}{4} +
\omega_b^2 \right \}^{1/2} - \frac{g_0^2}{2} \right ]  \; 
\exp \left ( - \frac{E_0}{ k_BT + D \kappa_0^2 } \right ) \; \; .
\end{equation}

\noindent
If we put the external noise intensity $D$ equal to zero, i.e., when 
external noise is absent, the above expression reduces to usual Kramers rate 
expression with $g_0^2 = \gamma$. We note here that $D \kappa_0^2/k_B$
defines a new {\it effective} temperature characteristic of the steady state
of the nonequilibrium open system. As expected this temperature is function
of the strength of external noise intensity $D$ and the coupling of the
external noise to the bath modes.

\section{ Conclusions 
\label{sd6}
}

Based on a system-reservoir microscopic model where the reservoir is 
modulated by an external, stationary and Gaussian noise with arbitrary
decaying correlation function, we have generalized the Kramers' theory
to calculate the steady state rate of escape from a metastable well. The main
conclusions of this study are as follows;

\noindent
(i) We have shown that since the reservoir is driven by the external noise
and the dissipative properties of the system depend on the reservoir, a
simple connection between the dissipation and the response function of the
medium due to the external noise can be established.

\noindent
(ii) This connection is important for realising the stationary state of the
thermodynamically open system characterized by an {\it effective} temperature
of the reservoir, which depends on the strength of the external noise.

\noindent
(iii) Provided the long time limit of the moments for the stochastic processes
pertaining to the external and internal noises characterized by arbitrary
decaying correlation functions exist, the expression for generalized
Kramers' rate of barrier crossing for the open system we derive here is
fairly general. The expression assumes simple forms in the specific limiting 
cases.

The creation of a typical nonequilibrium open situation by modulating a bath
with the help of
an external noise is not an uncommon phenomenon in applications and
industrial processing. The external agency generating noise does work on the
bath by stirring, pumping, agitating, etc., to which the system dissipates 
internally. In the present treatment we are concerned with a nonequilibrium
steady state characterized by an {\it effective} temperature which signifies 
a constant throughput of energy in contrast to thermal equilibrium defined
by an constant temperature. We believe that these considerations are likely
to be important in other related issues in nonequilibrium open systems.

\acknowledgments
SKB in indebted to Council of Scientific and Industrial Research (CSIR),
Government of India for financial support.


\begin{appendix}

\section{Calculation of variances}

The Laplace transform of $\gamma$ in Eq.(\ref{eq55}) is given 
by
\begin{eqnarray*}
\tilde{\gamma} (s) = \frac{ g_0^2 }{s \tau_c +1 } \; \; ,
\end{eqnarray*}

\noindent
and subsequently, we have
\begin{eqnarray*}
\tilde{h} (s) = \frac{ s\tau_c +1}{
\tau_c s^3 + s^2 + (\omega_0^2 \tau_c +g_0^2)s + \omega_0^2} \; \; .
\end{eqnarray*}

\noindent
For $\tau_c \neq 0$, the above expression can be simplified to
\begin{eqnarray*}
\tilde{h} (s) = \frac{ s+a }{
s^3 + a s^2 + b s + c_0}
\end{eqnarray*}

\noindent
where
\begin{eqnarray*}
a = \frac{1}{\tau_c} \; \; , \; \; b = \omega_0^2 + \frac{g_0^2}{\tau_c}
\; \; {\rm and} \; \;  c_0 = \frac{\omega_0^2 }{\tau_c} \; \; .
\end{eqnarray*}

\noindent
We define the following characteristic quantity
\begin{eqnarray*}
{\cal Q} \equiv -\frac{a^2b^2}{108} + \frac{b^3}{27} + \frac{a^3 c_0}{27}
- \frac{abc_0}{6} + \frac{c_0^2}{4}
\end{eqnarray*}

\noindent
and distinguish the cases : ${\cal Q} > 0$, ${\cal Q}=0$ and ${\cal Q} < 0$,
for which we have different forms of $h(t)$. For the case ${\cal Q}>0$
[ we do not give here the explicit expressions for the cases 
${\cal Q}=0$ and ${\cal Q} <0$ ] we find that the inverse
Laplace transform of $\tilde{h}(s)$ reads
\begin{equation}
\label{eqa1}
h (t) = c_1 e^{-\Delta_1 t} + c_2 e^{-\Delta_2 t} \; \sin(\beta t + \alpha)
\end{equation}

\noindent
where the coefficients $c_1$, $c_2$, $\Delta_1$, $\Delta_2$, $\beta$ and $\alpha$ are 
given by
\begin{mathletters}
\begin{eqnarray}
\Delta_1 & = & - {\cal A} - {\cal B} + \frac{a}{3} \; \; ,\\
\Delta_2 & = & \frac{1}{2}({\cal A} + {\cal B}) + \frac{a}{3} \; \; ,\\
\beta & = & \frac{\sqrt{3} }{2} ( {\cal A} - {\cal B} ) \; \; ,\\
c_1 & = & \frac{1}{ 2\Delta_2 - \Delta_1 - d } \; \; ,\\
d & = & \frac{ a (2\Delta_2 - \Delta_1) - \Delta_2^2 -\beta^2 }{ a - \Delta_1 } \; \; ,\\
{\cal A} & = & \left ( -\frac{a^3}{27} + \frac{ab}{6} - \frac{c_0}{2} + \sqrt{{\cal Q}}
\right )^{1/3} \; \; ,\\
{\cal B} & = & \left ( -\frac{a^3}{27} + \frac{ab}{6} - \frac{c_0}{2} - \sqrt{{\cal Q}}
\right )^{1/3} \; \; ,\\
c_2 & = & -\frac{c_1}{\beta} [ (d-\Delta_2)^2 + \beta^2 ]^{1/2} \; \; {\rm and}\\
\alpha & = & \tan^{-1} \left ( \frac{\beta}{d-\Delta_2} \right ) \; \; .
\end{eqnarray}
\end{mathletters}

\noindent
Here we note that for a physically allowed solution $\Delta_1$, $\Delta_2$
must be positive. Since by Eq.(\ref{eq19}) $h(t)$ depends on the memory
kernel $\gamma (t)$ which is of decaying type and all the moments, in
general, reach asymptotic constancy as shown in Sec.~{III}, these quantities
are positive (which depends on the correlation time $\tau_c$, the strength
of the noise and other potential parameters) which may be checked (after
some algebra) by considering the limiting cases such as $\tau_c \rightarrow 0$
and $\tau_c \rightarrow$ large.

Substituting Eq.(\ref{eq48}) into the expressions for variances [external
noise is $\delta$-correlated], namely into (\ref{eq25}) and (\ref{eq26})
we have after some lengthy algebra
\begin{eqnarray*}
\sigma_{xx}^2 (t) = \sigma_{xx}^{2(i)} (t) + \sigma_{xx}^{2(e)} (t)
\end{eqnarray*}

\noindent
where
\begin{eqnarray}
\sigma_{xx}^{2(i)} (t) & = & k_BT \left ( c_2 R + \frac{c_1}{\Delta_1} \right )
\left [ 2 - \omega_0^2 \left ( c_2 R + \frac{c_1}{\Delta_1} \right ) \right ]
\nonumber \\
& & + k_BT \left \{ -\frac{c_1}{\Delta_1} e^{-\Delta_1 t} \left [ 2 - 
2 \omega_0^2 c_2 R - \frac{2 \omega_0^2 c_1}{\Delta_1} + e^{-\Delta_1t}
\left ( \Delta_1 c_1 + \frac{\omega_0^2 c_1}{\Delta_1} \right ) \right ] \right.
\nonumber \\
& & - \frac{2 c_2 e^{-\Delta_2 t} }{\Delta_2^2 + \beta^2} \left [ 1 -
\omega_0^2 c_2 R + \frac{\omega_0^2 c_1}{\Delta_1} ( e^{-\Delta_1 t} - 1 )
\right ] \left [ \Delta_2 \sin (\beta t + \alpha) + \beta \cos (\beta t + \alpha)
\right ] \nonumber \\
& & - 2 c_1 c_2 e^{ - (\Delta_1+\Delta_2) t} \sin (\beta t + \alpha) \nonumber \\
& & - \frac{\Delta_2 \beta \omega_0^2 c_2^2 e^{-2\Delta_2t} }{ (\Delta_2^2+\beta^2)^2 }
\sin 2 (\beta t + \alpha)
- \frac{\beta^2 \omega_0^2 c_2^2 e^{-2\Delta_2t} }{ (\Delta_2^2+\beta^2)^2 }
\nonumber \\
& & \left. + \left [ 
\frac{ \omega_0^2 (2\beta^2 - \Delta_2^2) }{ (\Delta_2^2+\beta^2)^2 } -1 \right ]
c_2^2 e^{-2\Delta_2t} \sin^2 (\beta t + \alpha ) \right \}
\end{eqnarray}

\noindent
with
\begin{equation}
\label{eqa2}
R = \frac{1}{\Delta_2^2 + \beta^2} ( \Delta_2 \sin \alpha + \beta \cos \alpha)
\end{equation}

\noindent
and
\begin{eqnarray}
\label{eqa3}
\sigma_{xx}^{2(e)} (t) & = & 
2D \left ( \frac{\kappa_0}{g_0} \tau_c \right )^2
[ c_1^2 ( \omega_0^4 + \Delta_1^4 + 2\omega_0^2 \Delta_1^2 ) I_A (t) \nonumber \\
& & + c_2^2 \{ \omega_0^4 + (\Delta_2^2-\beta^2)^2 - 4\beta^2 \Delta_2^2 + 
2 \omega_0^2 ( \Delta_2^2 - \beta^2) \} I_B (t) \nonumber \\
& & + 2 c_1 c_2 \{ \omega_0^4 + \Delta_1^2 (\Delta_2^2-\beta^2)  
+ \omega_0^2 ( \Delta_1^2 + \Delta_2^2 - \beta^2) \} I_C (t) \nonumber \\
& & - 2 c_2^2 \beta \Delta_2 ( \Delta_2^2 -\beta^2 + \omega_0^2 ) I_D (t)
+ 4 c_2^2 \beta^2 \Delta_2^2 I_E (t) 
- 4 c_1 c_2 \beta \Delta_2 ( \Delta_1^2 + \omega_0^2 ) I_F (t) ] \; \; .
\end{eqnarray}

\noindent
Here the $I$'s are defined by
\begin{mathletters}
\begin{eqnarray}
\label{eqa4}
I_A (t) & = & \int_0^t e^{-2\Delta_1t} \; dt \; \; , \\
I_B (t) & = & \int_0^t e^{-2\Delta_2t} \sin^2 (\beta t +\alpha) \; dt \; \; , \\
I_C (t) & = & \int_0^t e^{-(\Delta_1+\Delta_2)t} \sin (\beta t +\alpha) \; dt \; \; , \\
I_D (t) & = & \int_0^t e^{-2\Delta_2t} \sin 2(\beta t +\alpha) \; dt \; \; , \\
I_E (t) & = & \int_0^t e^{-2\Delta_2t} \; dt \; \; {\rm and} \\
\label{eqa5}
I_F (t) & = & \int_0^t e^{-(\Delta_1+\Delta_2)t} \cos (\beta t +\alpha) \; dt \; \; .
\end{eqnarray}
\end{mathletters}

\noindent
Similarly
\begin{eqnarray*}
\sigma_{vv}^2 (t) = \sigma_{vv}^{2(i)} (t) + \sigma_{vv}^{2(e)} (t)
\end{eqnarray*}

\noindent
where
\begin{eqnarray}
\sigma_{vv}^{2(i)} (t) & = & k_BT - [ ( \Delta_1^2 + \omega_0^2 ) c_1^2 e^{-2\Delta_1t}
+ \beta^2 c_2^2 e^{-2\Delta_2t} \nonumber \\
& & - \beta \Delta_2 c_2^2 e^{-2\Delta_2t} \sin 2 (\beta t + \alpha ) +
( \Delta_2^2 + \omega_0^2 - \beta^2 ) c_2^2 e^{-2\Delta_2t}
\sin^2 (\beta t + \alpha )\nonumber \\
& & + e^{-(\Delta_1+\Delta_2)t} \{ 2c_1c_2 (\omega_0^2 + 
\Delta_1 \Delta_2 )\sin (\beta t + \alpha )
- 2 \Delta_1 \beta c_1 c_2 \cos (\beta t + \alpha ) \} ]
\end{eqnarray}

\noindent
and
\begin{eqnarray}
\label{eqa6}
\sigma_{vv}^{2(e)} (t) & = & 
2D \left ( \frac{\kappa_0}{g_0} \tau_c \right )^2
[ c_1^2 \Delta_1^2 ( \omega_0^2 + \Delta_1^2 )^2 I_A (t) \nonumber \\
& & + c_2^2 \{ ( \omega_0^2 + \Delta_2^2 - 3\beta^2)^2 \Delta_2^2 
- ( \omega_0^2 + 3 \Delta_2^2 - \beta^2)^2 \beta^2 \} I_B (t) \nonumber \\
& & + 2 c_1 c_2 \Delta_1 \Delta_2 ( \omega_0^2 + \Delta_1^2 )  
( \omega_0^2 - 3 \beta^2 + \Delta_2^2 ) I_C (t) \nonumber \\
& & - c_2^2 \beta \Delta_2 ( 3 \Delta_2^2 -\beta^2 + \omega_0^2 ) 
(\omega_0^2 - 3\beta^2 + \Delta_2^2) I_D (t) \nonumber \\
& & + c_2^2 \beta^2 ( \omega_0^2 -\beta^2 + 3 \Delta_2^2 ) I_E (t) \nonumber \\
& & - 2 c_1 c_2 \beta \Delta_1 ( \Delta_1^2 + \omega_0^2 ) 
( \omega_0^2 + 3 \Delta_2^2 -\beta^2 ) I_F (t) ]
\end{eqnarray}

\noindent
where, $I$'s are defined in Eq.(\ref{eqa4}-\ref{eqa5}). 
The explicit expression for 
$\sigma_{xv}^2 (t)$ can be derived from Eq.(\ref{eq27}).
From Eqs.(\ref{eq50}) and (\ref{eq52}), we calculate the stationary values
of the variances as follows,
\begin{eqnarray*}
\sigma_{xx}^2 (\infty) = \sigma_{xx}^{2(i)} (\infty) + 
\sigma_{xx}^{2(e)} (\infty)
\end{eqnarray*}

\noindent
where
\begin{equation}
\label{eqa7}
\sigma_{xx}^{2(i)} (\infty) = k_BT \left ( c_2 R + \frac{c_1}{\Delta_1} \right
) \left [ 2 - \omega_0^2 \left ( c_2 R + \frac{c_1}{\Delta_1} \right )
\right ]
\end{equation}

\noindent
and
\begin{eqnarray}
\label{eqa8}
\sigma_{xx}^{2(e)} ( \infty ) & = & 
2D \left ( \frac{\kappa_0}{g_0} \tau_c \right )^2
[ c_1^2 ( \omega_0^4 + \Delta_1^4 + 2\omega_0^2 \Delta_1^2 ) \frac{1}{2\Delta_1} 
+ c_2^2 \frac{1}{4\Delta_2} ( \Gamma^2 + 4 \beta^2 \Delta_2^2)  \nonumber \\
& & + \frac{c_2^2}{4 (\Delta_2^2 +\beta^2)} 
\{ \beta ( \Gamma^2 - 4\beta^2 - 4 \Delta_2^2 \beta ) \sin 2\alpha 
- \Delta_2 ( \Gamma^2 + 4\beta^2 \Gamma - 4 \Delta_2^2 \beta^2 ) \cos 2\alpha \}
\nonumber \\
& & + 2 c_1 c_2 \frac{\Delta_1^2+\omega_0^2}{ (\Delta_1+\Delta_2)^2 + \beta^2}
\nonumber\\
& & \times \; ( \{ \Gamma (\Delta_1+\Delta_2) + 2 \beta^2 \Delta_2 \} \sin \alpha + 
\beta \{ \Gamma - 2 \Delta_2 (\Delta_1+\Delta_2) \} \cos \alpha )  ]
\end{eqnarray}

\noindent
where
\begin{equation}
\Gamma = \omega_0^2 + \Delta_2^2 - \beta^2 \; \; .
\end{equation}

\noindent
Again substituting the values of $c_1$, $c_2$, $\Delta_1$, $\Delta_2$, $\alpha$,
$\beta$, $\Gamma$ and $R$ we obtain after a lengthy calculation the following 
result
\begin{equation}
\label{eqa9}
\sigma_{xx}^2 (\infty ) = \frac{k_BT + D \kappa_0^2}{\omega_0^2} \; \; .
\end{equation}

\noindent
Similarly
\begin{eqnarray}
\label{eqa10}
\sigma_{vv}^{2(e)} (\infty ) & = & 2D \left ( \frac{\kappa_0}{g_0} \tau_c 
\right )^2 \; [ c_1^2 \{ \frac{\Delta_1}{2} ( \omega_0^2 + \Delta_1^2 )^2 \}
\nonumber \\
& & + c_2^2 \frac{1}{4 \Delta_2} \{ \beta^2 ( \omega_0^2 + 3 \Delta_2^2 - \beta^2
)^2 + \Delta_2^2 ( \omega_0^2 +  \Delta_2^2 - 3 \beta^2 )^2 \} \nonumber \\
& & - \frac{c_2^2}{4 ( \Delta_2^2 + \beta^2 ) } \nonumber \\
& & \times \; ( \{ \beta \Delta_2^2 
( \omega_0^2 +  \Delta_2^2 - 3 \beta^2 ) ( \omega_0^2 +  5 \Delta_2^2 + \beta^2 )
+ \beta^3 ( \omega_0^2 +  3 \Delta_2^2 - \beta^2 )^2 \} \sin 2 \alpha
\nonumber \\
& & + \{ \Delta_2 \beta^2 ( \omega_0^2 +  3 \Delta_2^2 - \beta^2 )
( \omega_0^2 -  \Delta_2^2 - 5 \beta^2 ) + \Delta_2^3 ( 
\omega_0^2 +  \Delta_2^2 - 3 \beta^2 )^2 \} \cos 2 \alpha ) \nonumber \\
& & + 2 c_1 c_2 \frac{\Delta_1 (\Delta_1^2+\omega_0^2) }{ (\Delta_1+\Delta_2)^2 + \beta^2}
\nonumber \\
& & \times \; ( \{ \Delta_2 ( \Delta_1 + \Delta_2 ) ( \omega_0^2 +  \Delta_2^2 - 3 \beta^2 )
- \beta^2 ( \omega_0^2 +  3 \Delta_2^2 - \beta^2 ) \} \sin \alpha \nonumber \\
& & + \{ \beta \Delta_2 ( \omega_0^2 +  \Delta_2^2 - 3 \beta^2 ) - \beta
( \Delta_1 + \Delta_2 ) ( \omega_0^2 +  3 \Delta_2^2 - \beta^2 ) \} \cos \alpha ) ]
\; \; .
\end{eqnarray}

\noindent
Again putting the values of all the parameters we have,
\begin{equation}
\label{eqa11}
\sigma_{vv}^2 ( \infty ) = k_BT + D \kappa_0^2 \; \; .
\end{equation}

\noindent
Clearly,
\begin{equation}
\label{eqa12}
\sigma_{xv}^2 ( \infty ) = 0 \; \; .
\end{equation}

\noindent
The variances $\sigma_{xx}^2 (\infty)$, $\sigma_{vv}^2 (\infty)$ and 
$\sigma_{xv}^2 (\infty)$ yield $\phi (\infty)$ and $\psi (\infty)$ and other 
relevant quantities.

\end{appendix}

\end{document}